\documentclass[reprint,
superscriptaddress,
 amsmath,amssymb,
 aps, 
prb,
floatfix,
]{revtex4-2}
\usepackage{dcolumn}
\usepackage{bm}
\usepackage{mathtools}
\usepackage{graphicx}
\usepackage{color}
\usepackage{appendix}
\usepackage[colorlinks,bookmarks=false,citecolor=blue,linkcolor=black,urlcolor=blue]{hyperref}
\usepackage[nameinlink]{cleveref}
\Crefname{equation}{Equation}{Equations}
\Crefname{figure}{Figure}{Figures}
\Crefname{section}{Section}{Sections}
\Crefname{appendix}{Appendix}{Appendices} 
\Crefname{tabular}{Tabular}{Tabulars}
\crefname{equation}{Eq.}{Eqs.}
\crefname{eqnarray}{Eq.}{Eqs.}
\crefname{figure}{Fig.}{Figs.}
\crefname{section}{Sec.}{Secs.}
\crefname{appendix}{App.}{Apps.}
\usepackage{tikz}
\usetikzlibrary{positioning,shapes.geometric}
\usepackage{pgfplots}
\usepackage{SIunits}
\usepackage{pgfplotstable}
\usepgfplotslibrary{groupplots}
\pgfplotsset{width=7cm,compat=1.18}
\usepackage{placeins}
\usepackage{float}
\usepackage{lipsum}

\newcommand{\comm}[1]{{{#1}}}

\newcommand{\squareline}{\tikz{\node[rectangle,scale=0.7, draw=blue!100, fill=blue!100,  thick]{};\draw[-,blue] (-0.4,0) to (0.4,0) ; }}

\newcommand{\diamondline}{\tikz{\node[shape aspect=2,rotate=90,diamond,scale=0.4, draw=red!60, fill=white!5,  thick]{};\draw[-,red] (-0.4,0) to (0.4,0) ; }}

\newcommand{\opencircleline}{\tikz{\node[circle,scale=0.7, draw=black!60, fill=white!5,  thick]{};\draw[-] (-0.4,0) to (0.4,0) ; }}
\newcommand{\opensquare}{\tikz{\node[rectangle,scale=0.7, draw=blue!60, fill=white!5,  thick]{};  }}
\newcommand{\squaremark}{\tikz{\node[rectangle,scale=0.7, draw=blue!100, fill=blue!100,  thick]{}; }}
\newcommand{\diamondmark}{\tikz{\node[shape aspect=2,rotate=90,diamond,scale=0.4, draw=green!70!black, fill=green!70!black,  thick]{}; }}
\newcommand{\opendiamond}{\tikz{\node[shape aspect=2,rotate=90,diamond,scale=0.4, draw=green!70!black, fill=white!5,  thick]{}; }}
\newcommand{\opencircle}{\tikz{\node[circle,scale=0.7, draw=red!60, fill=white!5,  thick]{}; }}
\newcommand{\circlemark}{\tikz{\node[circle,scale=0.7, draw=red!60, fill=red!100,  thick]{}; }}
\begin{document}
\title{
Enhancement of superconducting stiffness in hybrid superconducting-metallic bilayers
}
\author{J. E. Ebot}
\affiliation{SUPA, Institute of Photonics and Quantum Sciences, Heriot-Watt University,
Edinburgh EH14 4AS, United Kingdom}
\author{Lorenzo Pizzino}
\affiliation{DQMP, University of Geneva, 24 Quai Ernest-Ansermet, 1211 Geneva, Switzerland}
\author{Sam Mardazad}
\affiliation{SUPA, Institute of Photonics and Quantum Sciences, Heriot-Watt University,
Edinburgh EH14 4AS, United Kingdom}
\author{Johannes S. Hofmann}
\affiliation{Max-Planck-Institut f\"ur Physik komplexer Systeme,N\"othnitzer Strasse 38, 01187 Dresden, Germany}
\author{Thierry Giamarchi}
\affiliation{DQMP, University of Geneva, 24 Quai Ernest-Ansermet, 1211 Geneva, Switzerland}
\author{Adrian Kantian}
\affiliation{SUPA, Institute of Photonics and Quantum Sciences, Heriot-Watt University,
Edinburgh EH14 4AS, United Kingdom}
\date{\today}

\begin{abstract}
    Boosting superconductivity by metallic reservoirs is the essence of Kivelson's bilayer proposal.
    One layer provides pairing to the electrons, while the weakly coupled metal provides additional phase coherence to those pairs by mediating extended-range pair-pair coupling.
    Demonstrating significant and unambiguous performance gains with strong-coupling methods for such set-ups had been difficult.
    In the present work, we study these systems doped away from half-filling, corresponding to a partially spin-polarized 1D Anderson- or Kondo-lattice.
    We show that this breaks the coexistence of dominant superconducting and density-density correlations decisively in favour or the former.
    Consequently, we provide evidence that in this doped regime, superconducting near-long-range order is not precluded by a small charge-gap in the thermodynamic limit, as we have recently shown to be the case at half-filling [J. E. Ebot \textit{et al.}, Phys. Rev. B \textbf{113}, 174530 (2026)]. 
    We study the complex manner in which the enhancement of superconductivity in the pairing layer depends on the parameters of the metal, and especially that both pairing-limited and stiffness-limited regimes may appear in these systems.
    In addition to superconducting bilayers, our results are relevant, via a particle-hole transformation, for heavy-fermion Kondo-lattice materials in magnetic fields, as we provide previously lacking insight on the competition between antiferromagnetic and easy-plane magnetism, as well as a route for comprehensive indirect tests of Kivelson's bilayer proposal well beyond previous capabilities.
\end{abstract}

\maketitle
\section{Introduction}
Superconducting systems appeared to be naturally caught between two conflicting aims~\cite{Emery1995}: 
maximising the amplitude of the order parameter, and maximising the stiffness of the phase of the paired electrons.
A strong pairing amplitude typically results in low pair-phase stiffness, which would limit superconducting performance (the ``stiffness-limited'' state).
Conversely, a high stiffness for the phase of the pairs often results in a weak pairing amplitude, which would then become the factor limiting overall superconducting performance (the ``amplitude-limited'' state).
Boosting superconductivity by means of sidestepping or at least minimizing this competition has gained much interest following Kivelson's proposal on enhancing the superconductivity in cuprate-layers~\cite{KIVELSON2002}. 
The essence of the proposal is that of a heterogeneous, composite system, where the pairing strength is optimized in one subsystem, and the superconducting stiffness is optimized in the other subsystem. 
The proposal originated from the theoretical work of~\cite{Emery1995} arguing that an optimal critical temperature $T_c$ for superconductivity is obtained in the crossover region between a phase-dominated and a pairing-dominated regime.
In practical terms, one can study and realize such heterogeneous systems, both for conventionally and unconventionally mediated pairing, by using a system that consists of a subsystem with some type of intrinsic pairing mechanism (``P-subsystem'' in the following), in contact with some type of metallic reservoir (``M-subsystem'', or just ``metal'')~\cite{KIVELSON2002,Erez2008,Yuli2008,Lobos2009,Gideon2012,Huang2017,JEEBOT2025,Zhang2025}.
These setups are most readily envisioned for one-dimensional~\cite{JEEBOT2025,JEEBOT2025_3} or two-dimensional~\cite{Erez2008,Gideon2012,Zhang2025} (1D or 2D) systems, i.e. where both subsystems have the same dimensionality.
But a 1D P-subsystem in contact with a (disordered) 2D metal has also been studied~\cite{Lobos2009}.
At the same time, the question as to how a version of Kivelson's proposal might be realized where both P- and M-subsystems are 3D, and their contact interface is not just a 2D one, is naturally hovering in the background.

In the field of reservoir-enhanced superconductivity that was thus established, theoretical work has encountered four interrelated challenges:

\textbf{(1)} How does one actually quantify and thus clearly demonstrate any enhancement of superconductivity in the P-subsystem, compared to the optimal cases that can be realized in the baseline of the isolated P-subsystem?
Meeting this challenge has turned out to be difficult.
Many-body numerics applied to this problem must be of a beyond-mean-field type due to the low-dimensional setting, which makes fluctuations too prominent for mean-field-based techniques. 
This entails computational resource-requirements that scale significantly with system size.
It is thus costly to look for regimes where superconducting properties are boosted significantly by the metal, as this requires attacking a fairly high-dimensional parameter-space, with little analytical guidance available, as discussed in item (3) below.
Work on 2D systems has shown that Kivelson's proposal can impart superconducting stiffness when none is available in the isolated P-subsystem, because it was modelled as a collection of disconnected negative-$U$ centers~\cite{Erez2008,Gideon2012}. 
The metallic layer then allowed the overall system to establish a superconducting quasi-order below a finite Berezinskii-Kosterlitz-Thouless (BKT) temperature, $T_{\rm BKT}$.
But so far, there has been no version of Kivelson's proposal in 2D that evidences a $T_{\rm BKT}$ better than that of the isolated P-subsystem with optimal parameters~\cite{Fontenele2022}.
However, recent results provide some examples of P-subsystems in a stiffness-limited state (i.e. in the opposite of the amplitude-limited state) for which metallic reservoirs can boost $T_{\rm BKT}$ somewhat beyond the optimal value for a comparable but isolated P-subsystem~\cite{Zhang2025,JEEBOT2026}.

\textbf{(2)} These latest findings on the 2D version of the Kivelson proposal still come with uncertainties due to the limited system sizes that can be simulated.
This restricts the accuracy of any extrapolation to the thermodynamic limit.
These problems become more pronounced when trying to move from P-subsystems in a stiffness-limited state (and thus with short coherence lengths) towards ones that are amplitude-limited (i.e. where coherence length could rapidly outgrow numerically feasible system sizes).
Our own study of the 1D version of Kivelson's proposal~\cite{JEEBOT2025} had in part been motivated by the fact that far larger linear sizes and thus the thermodynamic limit can be reached with modern numerics based either on matrix product states (MPS)~\cite{Schollwock2011} or on auxiliary field quantum Monte-Carlo (AFQMC)~\cite{10.21468/SciPostPhysCodeb.1-v2.4,10.21468/SciPostPhysCodeb.1-r2.4}.
This allowed us to show unambiguously that 1D systems comprised of a pairing and a metallic chain can far outperform either isolated chains or homogeneous ladders with intrinsic pairing, both inside and outside the stiffness-limited regime.
Our approach permitted insight, previously lacking, into how the P- and M-subsystems modify each other's properties, and how this can lead to significant enhancements of superconducting properties. 
But these results are still all based on numerics.
What remains elusive is quantitative analytical insight and thus the ability to predict - as opposed to painstakingly trying - how traversing the high-dimensional space of microscopic parameters impacts any enhancements of superconductivity.

\textbf{(3)} In light of the numerical effort required to study Kivelson's proposal, especially in 2D and even more so in 3D, complementary analytical treatments with their ability to address the thermodynamic limit are sorely needed.
Prior work~\cite{Erez2008,Lobos2009,Gideon2012}, has yielded important insight into how weak coupling between P- and M-subsystem can mediate long-ranged pair-pair coupling within the P-subsystem, which substantially strengthens the superconducting properties.
At the same time, these approaches have so far invariably been based on low-order perturbation theory in the coupling between the two subsystems.
These approximations left the properties of the metal practically unchanged, and predicted long-range mediated pair-pair coupling~\cite{Erez2008,Lobos2009,Gideon2012} that should be able to drive even 1D P-subsystems to true superconducting long-range order - in apparent contradiction to the Mermin-Wagner theorem (MWT)~\cite{Thierrybook2003}.
Our recent numerical work~\cite{JEEBOT2025,JEEBOT2025_3} has resolved this apparent paradox.
It demonstrated that the pairing subsystem induces a gap in the metal via the proximity effect.
In turn, this curtails the range of pair-pair-coupling mediated by the metal to something that is still extended, but effectively finite-ranged.
Thus, to a degree, the P-subsystem is ``poisoning'' the M-subsystem that is supposed to supply additional pair-phase-stiffness.
But our data shows that even the metal thus modified can still yield enhancements to superconducting susceptibilities and pair-pair correlations.
These enhanced correlations, while not corresponding to true long-range order, can still approach such ordering very closely, far in excess of what any isolated P-subsystem would be capable of in 1D.

\begin{figure}
  \centering
\includegraphics[scale=1]{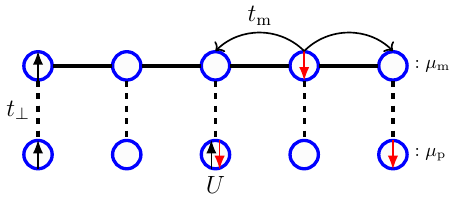}
\caption{(a) Schematic representation of the Hamiltonian in Eq. (\ref{model_equ}).
It comprises of a pairing subsystem (``P-subsystem'') with on-site pairing $U$ (and chemical potential $\mu_{\rm m}$, when performing AFQMC simulations, which are in the grand-canonical regime), and a metallic subsystem (M-subsystem; we will use `metal' interchangeably) with nearest-neighbour tunneling $t_{\rm m}$ and chemical potential term $\mu_{\rm m}$.
Coupling between the subsystems is provided by tunneling with amplitude $t_\perp$.}
\label{pictorial}
\end{figure}

\textbf{(4)} Viewed in summary, challenges (1) to (3) suggest that in order to arrive at a version of Kivelson's proposal that is tractable with non-perturbative analytics and that can incorporate the back-action of the P-subsystem onto the metal, the model should be simplified as much as possible.
We have followed such a strategy in ref.~\cite{JEEBOT2025_3}.
There, we considered two chains, one composed solely of disconnected negative-$U$ centers, the other of a non-interacting metal, that are weakly coupled to each other (c.f.~\cref{pictorial}), and which both sit at half-filling, i.e. at $n=1$ in both chains.
We showed that this system can be mapped to the periodic 1D Anderson- and Kondo-lattices.
We thereby established a new connection, between Kivelson's proposal on the enhancement of superconducting properties on the one hand, and the mechanism by which the itinerant electrons in the conduction band seek to drive the spins within the lattice of localized orbitals towards antiferromagnetic (AFM) long-range order inside the heavy fermion materials (HFMs)~\cite{Stewart1984,Das2014,Wang2017,Fumega2024} on the other hand.
While we derived this link in general, in lattices of arbitrary dimensionality, in ref.~\cite{JEEBOT2025_3} we supplied complementary analytical and numerical approaches specifically for the 1D case.
We showed analytically that this version of Kivelson's proposal exhibits an induced spin-gap inside the metal at any coupling to the P-subsystem.
At the same time, we also found that due to being at half-filling there was also always a charge gap in these systems, at any coupling between the P- and M-subsystems.
While this gap is exponentially small at small coupling, its existence will ensure that the degenerate near-long-range superconducting and change-density correlation will be transitory phenomena, and will ultimately be suppressed in the thermodynamic limit.
We find this degeneracy to be fulfilled exactly even at the largest possible scales that can be simulated.
Translated to periodic Kondo- and Anderson lattices at half-filling, our results mean that these systems can appear to show AFM near-long-range order over large scales, but that they will ultimately still be paramagnets.
This translation maps the superconducting correlations of Kivelson's proposal to the AFM spin-correlations in the $x$- and $y$-directions for the Anderson-/Kondo-lattices, while the density-density correlations map to spin-spin-correlations in the $z$-direction.

The present work thus builds on that in ref.~\cite{JEEBOT2025_3} by considering the 1D version of Kivelson's proposal shown in \cref{pictorial}, but now doped away from half-filling, $n\neq 1$.
We show that moving away from half-filling, superconducting correlations can still exhibit near-long-range behaviour in these systems, while density-density correlations cease to be degenerate with these and now decay much faster.
We also provide comprehensive data that the charge-gap, which ultimately suppresses any correlations in the infinite-size limit for the half-filled system, is almost certainly absent in the doped systems.
Our results further illustrate the complex way in which the metal enhances the superconducting properties of the P-subsystem, based on the strength of the coupling between the two subsystems:
As we found to be the case in~\cite{JEEBOT2025,JEEBOT2025_3} for related systems, here the proximity effect fundamentally changes the single-electron propagator of the metal from long-ranged (i.e. algebraically decaying) to short-ranged (exponentially decaying) via induced spin-gap.
But even with this modification, the metal can mediate pair-pair-coupling within the P-subsystem over extended distances, which can yield exceptionally high superconducting susceptibilities.
We find that as the effective range of this mediated coupling is increased, the superconducting susceptibility of the P-subsystem may either improve further (pair-phase-limited regime), or be somewhat degraded (amplitude-limited regime).
Likewise, we show that changing the coupling between the P- and M-subsystems allows to tune between regimes where density imbalance between the layers further enhances the superconducting properties of the P-subsystem, and those where imbalance lessens the enhancement.
Our results can thus serve as future benchmarks for adapting the analytical approaches that we used in ref.~\cite{JEEBOT2025_3} for half-filled systems to the doped case, which is more challenging and will require separate work.

Mapped to the Anderson-/Kondo-lattices, doping away from half-filling corresponds to spin-polarizing the system, e.g. via an external magnetic field.
We thereby predict that spin-polarizing materials that realize such lattices, as many HFMs are thought to do, will result in a shift from dominant AFM correlations to dominant easy-plane magnetic correlations.
Our results imply that the HFMs exposed to magnetic fields may be used as indirect test-beds for Kivelson's bilayer proposal. 
This would include the otherwise hard-to-realize case of P- and M-subsystems that are both 3D, and that are coupled to each other across their entire respective volumes.
Our results on the modification of the metal-properties by the proximity effect map to the Anderson-/Kondo-lattices. 
We thereby confirm that the back-action of the dense localized-spin subsystem onto the conduction band leads to Ruderman-Kittel-Kasuya-Yosida (RKKY) interactions that no longer decay algebraically (as predicted from analytical calculations for dilute localized spins~\cite{Rusin2017,Nejati2017}), but exponentially instead.
We had previously shown the same at half-filling~\cite{JEEBOT2025_3}. 

This article is structured in the following way: \cref{sec:setup} introduces the system Hamiltonian and the observables, explains the mapping to the periodic Anderson- and Kondo-lattices and provides an overview of the different parameter-regimes that we have studied; \cref{sec:res} then contains the main numeric results; in \cref{sec:impl} we discuss the implications of our findings for the periodic Anderson- and Kondo-lattices, and \cref{sec:disc} finally provides a summary and outlook for future directions.

\section{Hamiltonian, observables and parameter regimes}
\label{sec:setup}
We considered the lattice Hamiltonian depicted in \cref{pictorial} doped away from half-filling:
\begin{align}
\label{model_equ}
    \hat{H}&=-t_{\rm m}\sum_{i,j,\sigma}^{} (\hat{c}_{i,{\rm m},\sigma}^{\dagger}\hat{c}_{j,{\rm m},\sigma} + {\rm h.c.})-\mu_{\lambda}\sum_{i,\lambda,\sigma}^{} \hat{n}_{i,\lambda,\sigma} &\nonumber\\
    &-{t}_\perp \sum_{i,\sigma}^{}(\hat{c}_{i,{\rm p},\sigma}^{\dagger}\hat{c}_{i,{\rm m},\sigma}+ {\rm h.c.}) - U\sum_{i}^{}\hat{n}_{i,{\rm p},\uparrow}\hat{n}_{i,{\rm p},\downarrow}.&
\end{align}
%
Here, $c_{j,\lambda,\sigma}$ ($c^\dagger_{j,\lambda,\sigma}$) are fermionic annihilation (creation) operators for an electron on site $j$ with spin $\sigma$ $\in \{\uparrow\downarrow\}$ in chain $\lambda$, and ${n_{j,\lambda,\sigma}=c^\dagger_{j,\lambda,\sigma}c_{j,\lambda,\sigma}}$ are the derived density-operators.
The index $\lambda = {\rm p}$ or ${\rm m}$ labels the disconnected negative-$U$ sites in the pairing chain, the P-subsystem, and the sites in the metallic chain, the M-subsystem, respectively. 
Likewise, these labels pertain to Hamiltonian and other parameters of the chains:
$t_{\rm m}$ denotes the nearest-neighbour tunneling inside the M-subsystem, while the chemical potential $\mu _{\rm m}$ adjusts its density $n_{\rm m}$;
when working in the grand-canonical regime, as we do for the AFMQC-based calculations, we also include the chemical potential $\mu _{\rm p}$ to independently adjust the density $n_{\rm p}$ of the P-subsystem;
conversely, in the canonical regime used for calculations using the density matrix renormalization group (DMRG)~\cite{Schollwock2011}, we instead fix both $n_{\rm p}$ and $n_{\rm m}$ by setting $\mu _{\rm m}$ together with the global numbers of spin-up electrons $N_\uparrow$ and spin-down electrons $N_\downarrow$.
The strength of on-site pairing in the P-subsystem is parametrized via the on-site interaction $U>0$.
The pairing sites are coupled to the sites of the M-subsystem through a single-particle tunneling term with amplitude $t_\perp$.
In the present work we are interested mostly, but not exclusively, in the weakly coupled regime in which single-electron tunneling between the subsystems is largely suppressed and the leading process consists of electron-electron or electron-hole exchanges  in-between the P- and M-subsystem.
This is generally achieved by sticking to the regime ${t_\perp  < E_{\rm p}}$, where $E_{\rm p}$ is the pairing gap~\cite{Gunnar2023,JEEBOT2025,JEEBOT2025_3}. 
In the present system, this gap is given by ${E_{\rm p} = U}$.

Performing a selective particle-hole transformation as in \cite{JEEBOT2025_3}, the Hamiltonian in \cref{model_equ} maps to the periodic Anderson model:
\begin{align}\label{al_ham}
    \hat{H}_{\rm AL} =& -t_{\rm m}\sum_{i,\sigma}^{}(\hat{d}_{i,m,\sigma}^{\dagger}\hat{d}_{i+1,m,\sigma} + {\rm h.c.})&\nonumber \\
    +& U\sum_{i}(\hat{n}_{i,p,\uparrow}-\frac{1}{2})(\hat{n}_{i,p,\downarrow}-\frac{1}{2}) & \\
    -& {t}_{\perp}\sum_{i,\sigma}^{}(\hat{d}_{i,p,\sigma}^{\dagger}\hat{d}_{i,m,\sigma} + {\rm h.c.}) -\mu_{{\rm m}}\sum_{i}(\hat{n}_{i,{\rm m},\uparrow} - \hat{n}_{i,{\rm m},\downarrow})&
    \nonumber\\
    -&\frac{U}{2}\sum_{i}(\hat{n}_{i,p,\uparrow} - \hat{n}_{i,p,\downarrow}+\frac{1}{2}) -\mu_{{\rm p}}\sum_{i}(\hat{n}_{i,{\rm p},\uparrow} - \hat{n}_{i,{\rm p},\downarrow})&. \nonumber
    \nonumber
\end{align}
Here, the chemical potentials $\mu_{\lambda}$ takes on the meaning of external magnetic fields that adjust the spin imbalances within the two chains, and $U$ is now an onsite Coulomb repulsion.
As also discussed in ref.~\cite{JEEBOT2025_3} and references therein, in the weak-coupling limit $t_\perp/U\ll 1$ this model can be mapped to the periodic Kondo-lattice, though we will not draw on this mapping here.

We have numerically computed the $s$-wave pair-pair correlations $C_{\lambda }(i,j) $ and the metallic single-particle correlations $S_{\rm m}(i,j)$ for \cref{model_equ} at both zero and finite temperature, using DMRG and AFQMC respectively, where
\begin{align}
    \label{eq:corr_functions1}C_{\lambda}(i,j) &=\left\langle  \hat{c}^{\dagger}_{i,\uparrow,\lambda} \hat{c}^\dagger_{i,\downarrow,\lambda} \hat{c}_{j,\downarrow, \lambda } \hat{c}_{j,\uparrow, \lambda } \right\rangle , 
    \\
    \label{eq:corr_functions2}S_{\rm m}(i,j) &= \left\langle  \hat{c}^{\dagger}_{i,\sigma,\rm m}  \hat{c}_{j,\sigma,\rm m} \right\rangle.
\end{align}
At zero temperature, we also compute the density-density correlations:
\begin{align}
\label{eq:corrs_3} N_{\lambda}(i,j) &= \left\langle  (\hat{n}_{i,\lambda}-\langle \hat{n}_{i,\lambda} \rangle )(\hat{n}_{j,\lambda}- \langle \hat{n}_{j,\lambda} \rangle) \right\rangle,
\end{align}
where ${\hat{n}_{i,\lambda}:=\hat{n}_{i,\lambda,\uparrow}+\hat{n}_{i,\lambda,\downarrow}}$.
Due to the MWT, there is never any long-range order in these 1D systems, so these correlation functions will decay either algebraically, or even faster than that.
Specifically, the spatial dependency of the pair-pair correlation function in \cref{eq:corr_functions1}) is governed by  
\begin{align}
    C_{\lambda} (i,j) &=& {\left[ \frac{\pi A_{C,\lambda}(\beta)}{\left( \xi_{C,\lambda}(\beta) \sinh(\pi |i-j|/\xi_{C,\lambda}(\beta))\right ) } \right]^{K_{\lambda}^{-1}}}, 
 \label{eq:c_fitting_function}
\end{align}
as derived from conformal-field theory for the generic physics of ungapped charge-modes~\cite{Thierrybook2003}, which our data indicates both subsystems realize here.
The Tomonaga-Luttinger-liquid (TLL) parameter $K_{\lambda}$, also controls the scaling behavior of the pair-pair susceptibilities, which diverge algebraically with the power ${K_{\lambda}^{-1}-2}$ as the temperature is lowered.
The thermal correlation length $\xi_{C,\lambda}$ is equal to the speed of sound times the inverse temperature $\beta =1/T$ for a generic isolated chain.
For these hybrid systems however, $\xi_{C,\lambda}$ is both greatly enhanced over the isolated baseline and scales superlinearly in $\beta$~\cite{JEEBOT2025}, as extracted from fitting \cref{eq:c_fitting_function} to our numerical data, which also yields the non-universal amplitude $A_{C,\lambda}$.
The single-particle correlations (at $T=0$) in \cref{eq:corr_functions2} can be fitted by 
\begin{align}
 S_{\rm m}(i,j) =  {A_{S,\rm m}e^{-|i-j|/\xi_{S,\rm m}}}.
 \label{eq:s_fitting_function}
\end{align}
The exponential decay, which is controlled by the length-scale $\xi_{S,\rm m}$, is the direct expression of the gap induced in the metal by the coupling to the P-subsystem, i.e. a manifestation of the proximity effect.
Further details of our data analysis are provided in \cref{app:DMRG}.

Under the spin-selective particle-hole mapping that yields the Anderson-lattice Hamiltonian, \cref{al_ham}, from the original Hamiltonian, \cref{model_equ}, the superconducting and density-density correlations map to spin-spin correlation functions in the $x-y$- and $z$-directions respectively~\cite{JEEBOT2025_3}:
\begin{equation}
    \begin{aligned}\label{eq:corr_trafo}
        C_{\lambda} (i,j) &\rightarrow (-1)^{i-j}\langle \hat{S}^+_{i,\lambda} \hat{S}^{-}_{j,\lambda} \rangle \\
        N_{\lambda}(i,j) &\rightarrow 4\langle (\hat{S}^z_{i,\lambda} - \langle \hat{S}^z_{i,\lambda}\rangle) ( \hat{S}^z_{j,\lambda} - \langle \hat{S}^z_{j,\lambda} )\rangle.
    \end{aligned}
\end{equation}

In the present work we have studied three parameter regimes:
\begin{eqnarray}\nonumber
   && (1) \quad 1 \leq U \leq 4 , t_{\perp} =0.4, t_{\rm m}=1.0, \\
\nonumber&&(2)\quad U=10, 7 \leq t_{\rm m} \leq 10, 1 \leq t_{\perp} \leq 3,\\
\nonumber&&(3)\quad U=4, 1.0 \leq t_{\perp}  \leq 3.0, t_{\rm m}=1.0,
\end{eqnarray}
In regime~1 and regime~3, we have considered three different cases of incommensurate densities, motivated by our original work~\cite{JEEBOT2025}.
These are the diluted case ($n_{\rm p} < n_{\rm m}$), the non-diluted case ($n_{\rm p} >  n_{\rm m}$), and the case of nesting, wherein the nominal Fermi wave-vectors $k^\lambda_{\rm F}=\pi n_\lambda/2$ are close to each other ($n_{\rm p} \approx n_{\rm m}$). 
Concretely, and inspired by parameter sets first generated in our prior work, we have chosen ($n_{\rm p} = 0.67$, $n_{\rm m} = 0.49$); ($n_{\rm p} = 0.51$, $n_{\rm m} = 0.65$); ($n_{\rm p} = 0.39$, $n_{\rm m} = 0.77$). 
For regime~2, we have studied two diluted cases, $n_{\rm p}=0.2$, $n_{\rm m}=0.8$ and $n_{\rm p}=0.3$, $n_{\rm m}=0.7$, and one case of nesting, $n_{\rm p}=0.42$, $n_{\rm m}=0.58$, again driven by equivalent cases first considered in~\cite{JEEBOT2025}.
In regime~1 and regime~3 we target a global density of electrons $n=0.58$, and of $n=0.5$ in regime 2.  
We note that the three regimes seem to have no common unit of energy.
This has been intentional.
While we could have chosen e.g. either $U$ or $t_{\rm m}$ to act as a unit, this would significantly complicate direct comparisons with the systems that we first studied in ref.~\cite{JEEBOT2025}.
These had been special cases of regime~1 and regime~2 in the present work except for the lack of a kinetic energy term in the P-subsystem, which ref.~\cite{JEEBOT2025} did incorporate.

\section{Results}\label{sec:res}

\begin{figure}[t!]
    \centering
    \includegraphics[scale=1]{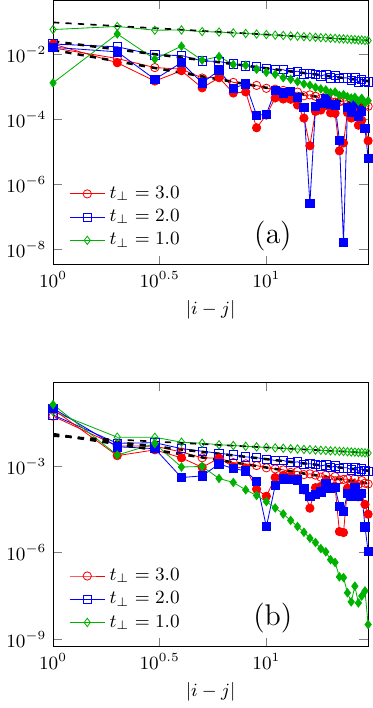}
    \caption{
     $C_{\lambda}(i,j)$ (open markers) and $|N_{\lambda}(i,j)|$ (shaded markers) of \textbf{(a)} P-subsystem and \textbf{(b)} M-subsystem in regime $3$ with $L=100$, $n_{\rm p}=0.38$ and $n_{\rm m}=0.78$. 
    $C_{\rm \lambda}(i,j)$  has infinite  correlation length and $|N_{\rm p}(i,j)|$ is suppressed. 
    %
    %
     $C_{\lambda}(i,j)$ (open markers:  \protect\opendiamond,\protect\opensquare,\protect\opencircle) and $|N_{\rm m}(i,j)|$ (shaded markers: \protect\diamondmark,\protect\squaremark,\protect\circlemark).
    Dashed lines are the power-law fit of $C_{\lambda}$, with fitting parameters $K^{-1}_{\rm p}$ and $A_{\rm C, p}$.
    }
    \label{fig:nocross_over}
\end{figure}
\begin{figure}[t!]
    \centering
    \includegraphics[scale=1]{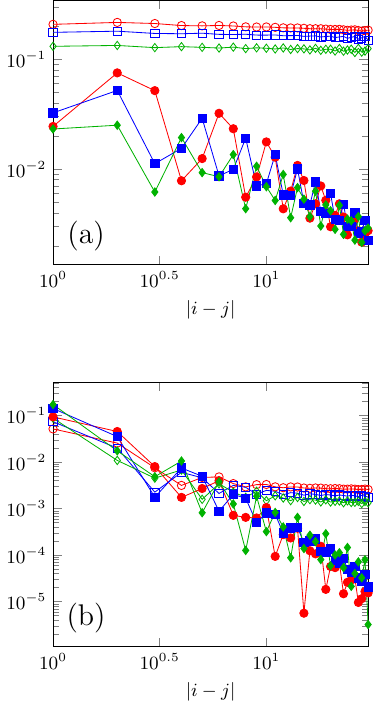}
    \caption{
     $C_{\lambda}(i,j)$ (open markers, \protect\opendiamond,\protect\opensquare,\protect\opencircle) and $|N_{\lambda}(i,j)|$ (shaded markers, \protect\diamondmark,\protect\squaremark,\protect\circlemark) in regime $1$ at $U=4$ and $L=100$. 
    (a) P-subsystem and (b) M-subsystem.
    $n_{\rm p}=0.67, n_{\rm m}=0.49$ (circle), $n_{\rm p}=0.51, n_{\rm m}=0.65$ (square) , $n_{\rm p}=0.39, n_{\rm m}=0.77$ (diamond).
    }
    \label{fig:U4pair-pair_corr}
\end{figure}

When we had previously studied Hamiltonian~\cref{model_equ} at half-filling (i.e. at $n=1$ in both subsystems), we had found that this system does not just have a global spin-gap due to the P-subsystem.
It also always exhibits a global charge gap as well, at any non-zero-value of $t_\perp$~\cite{JEEBOT2025_3}.
We showed that this gap is exponentially small when $t_\perp\ll U$, resulting in these systems appearing to possess degenerate near-long-range order in the superconducting and density-density channels across large scales.
Depending on the exact value of $t_\perp/U$, these scales can easily become larger than those that can be simulated at present.

Numerically, the marker of the charge gap at $n=1$ has been the rapid appearance of crossover behaviour from (near-long-ranged) algebraic decay of $C_\lambda$ and $N_\lambda$ to exponential decay as $t_\perp$ has been raised.
In the absence of the analytical theory that we developed for $n=1$, which remains to be adapted for the doped systems we study here, we rely on numerics to investigate whether the charge gap persists here.
Regime~3  in the present work is specifically meant to do just that.
At $n=1$ we had seen a crossover happening for a systems with $L=100$ rapidly beyond $t_\perp=0.55$ for $U=4.0$ and $t_{\rm m}=1.0$. 
Once doped, the system exhibits the behaviour summarized in \cref{fig:nocross_over}:
while the overall amplitude of $C_\lambda$ does decrease as $t_\perp$ is increased, which is to be expected, there is no indication that either $C_\lambda$ or $N_\lambda$ measurably deviate from algebraic scaling, even when $t_\perp$ rises up to values of $3.0$.
This data also provides an example for another change from the undoped regime that we have consistently found across all calculations in this work, further illustrated for regime~1 at $U=4.0$ in \cref{fig:U4pair-pair_corr}.
Namely, the degeneracy between superconducting and density-density correlations is  broken strongly in favour of the former.
As exemplified in \cref{fig:U4pair-pair_corr} this applies both to the P- as well as to the M-subsystem.

\begin{figure}[t!]
\centering
\includegraphics[scale=1]{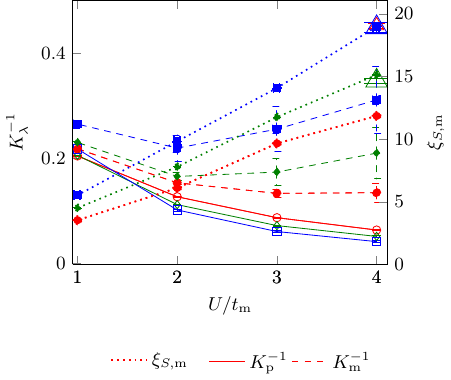}
    \caption{
    The Luttinger liquid parameters ($K_{\rm p}^{-1}$ and $K_{\rm m}^{-1}$) and single particle length scale $\xi _{\rm m}$ in regime~1 versus $U$ with $L=200,t_{\perp}=0.4$ 
    $n_{\rm p}=0.38 \pm 0.01,n_{\rm m}=0.78 \pm 0.01$ (\protect\squaremark, \protect\opensquare ), $n_{\rm p}=0.51,n_{\rm m}=0.65$ (\protect\diamondmark, \protect\opendiamond ),
     $n_{\rm p}=0.67\pm 0.01,n_{\rm m}=0.49\pm 0.01$ (\protect\circlemark, \protect\opencircle). 
    The markers~$\triangle$ are $K^{-1}_{p}$ for the Hubbard ladder in Ref~\cite{JEEBOT2025}.
    }
    \label{U4regime}
\end{figure}

\comm{With the presence of a charge gap in these doped systems thus being highly unlikely, we proceed on the assumption that the critical behaviour that we observe in both $C_\lambda$ and $N_\lambda$ reflects the true bulk physics.}
It is therefore not a transitory intermediate-scale effect as we found it to be in the undoped regime.
As the superconducting correlations $C_\lambda$ consistently dominate the density-density ones, in the following we are focusing on the TLL-parameters $K_{\lambda}$ that characterize their power-law decay with distance $|i-j|$ at zero temperature, $T^{-1}=\beta\rightarrow\infty$ (c.f. \cref{eq:c_fitting_function}).
We had previously shown in ref.~\cite{JEEBOT2025} that when the isolated pairing subsystem would be in a stiffness-limited state, upon coupling to the 1D metal the TLL-parameter $K_{\rm p}^{-1}$ is markedly improved (i.e. lowered) as the single-particle length-scale $\xi_{S,{\rm m}}$ is  growing (cf. \cref{eq:corr_functions2} and \cref{eq:c_fitting_function}). 
We recall that $K_{\rm p}^{-1}$ controls both the power-law decay of the pair-pair-correlations $C_{\rm p}$ and the superconducting susceptibility, while $\xi_{S,{\rm m}}$ characterizes the single-particle Green's function in the metal.
Conversely, we had also seen that when the isolated P-subsystem is in a state where the pairing amplitude limits its performance instead, coupling to the metal the results in $K_{\rm p}^{-1}$ being minimal when $\xi_{S,{\rm m}}$ is as low as possible, while always being well improved relative to the best isolated case, $K_{\rm p}^{-1}=1/2$. 
This minimizes the reverse proximity effect by which the metal weakens the strength of pairing inside the P-subsystem.
We therefore examine and plot $K_{\rm p}^{-1}$ and $\xi_{S,{\rm m}}$ together in multiple figures in this work.
To these, we now add the TLL-parameter $K_{\rm m}^{-1}$ characterizing the superconducting properties that the P-subsystem induces in the metal, starting with the data shown in \cref{U4regime} for regime~1.
First of all, at $U=4.0$, this data reveals a profound difference to the analogous case that we had studied in ref.~\cite{JEEBOT2025} for a modified system in which we had included direct nearest-neighbour tunneling, $t_{\rm p}=1.0$, between the negative-$U$ sites of the P-subsystem.
With this modification, the isolated P-subsystem is marked by a superconducting stiffness $\rho_s$ that is close to the maximally possible one~\cite{Thierrybook2003}.
While that system still profits from being coupled to the metallic subsystem, achieving $K_{\rm p}^{-1}$-values that would be out of reach in its isolated state, we also found it to be in the amplitude-limited regime.
We ascertained this by observing that in order to achieve the lowest possible $K_{\rm p}^{-1}$-values, this system would have to be tuned to be very close to nesting, i.e. matching densities in both subsystems (c.f. Fig.~2a in~\cite{JEEBOT2025}).
This nesting maximised the proximity-effect experienced by the metal, which in turn resulted in the lowest values of $\xi_{S,{\rm m}}$ and therefore yielded the lowest possible impact of the reverse proximity effect onto the pairing subsystem.
In short, in ref.~\cite{JEEBOT2025} nesting resulted in only minimally weakening the intrinsic pairing of the P-subsystem.
At the same time, the metal could still mediate some extended-range pair-pair coupling within the P-subsystem, wherein electron-pairs enter the metal, propagate separately but coherently up to distances of order $\xi_{S,{\rm m}}$, and then re-enter the P-subsystem as a pair.
This lowers $K_{\rm p}^{-1}$ below the minimum possible for the isolated limit at $t_\perp=0$.

The data for the case of $U=4.0$ in regime~1 shown in \cref{U4regime} exhibits a drastic change to these previously found behaviors.
Without the direct tunneling between the negative-$U$ sites that is present in ref.~\cite{JEEBOT2025}, $K_{\rm p}^{-1}$ is lowered, and thus improved, by factors of about $6.7$ to $10.7$ compared to the values that we had found in ref.~\cite{JEEBOT2025} at $U=4.0$.
We also indicate those $K_{\rm p}^{-1}$-values that the systems in ref.~\cite{JEEBOT2025} achieved in \cref{U4regime}).
The magnitude of this effect is almost independent of the values of the three $(n_{\rm p},n_{\rm m})$-pairs that we studied in the present work.
As we have removed any intrinsic kinetic term proportional to $t_{\rm p}$ from the P-subsystem in the present work, the pairing gap $E_{\rm p}$, which had been approximately $1.5t_{\rm m}$ in the related systems from ref.~\cite{JEEBOT2025} where we had set $t_{\rm m}=t_{\rm p}$, shoots up to be $E_{\rm p}=U=4.0t_{\rm m}$ for the rightmost set of data in \cref{U4regime}.
Coupled with the lack of any intrinsic kinetic energy, this pushes the P-subsystem firmly from the amplitude- to the stiffness-limited regime.
This explains why the P-subsystem now profits to a much greater extent from values of the single-particle length $\xi_{S,{\rm m}}$ that are comparable or slightly below those that we had found at comparable density-pairs $(n_{\rm p},n_{\rm m})$ in ref.~\cite{JEEBOT2025}.
This behaviour of $\xi_{S,{\rm m}}$ indicates that the removal of the kinetic energy-term from the P-subsystem slightly enhances the proximity effect experienced by the metal:
as the effective pairing energy in the P-subsystem is strongly boosted, the induced pairing experienced by the metal is (modestly) enhanced.
The phase-limited nature of this regime is further evidenced by the lowering of $K_{\rm p}^{-1}$ as the density-imbalance between M- and P-subsystems is increased.
This matches the behaviour that we had previously found for phase-limited systems (c.f. Fig.~2b in ref.~\cite{JEEBOT2025}).
Even though this effect is slight given that the values of $K_{\rm p}^{-1}$ are already very low, it exceeds the error bars, and the data in \cref{U4regime} shows it to be consistent across a range of $U$-values, as discussed next.

What is particularly striking is the change brought on by lowering $U$ in this regime for any of the three densities, also shown in \cref{U4regime}, while $t_\perp$ is kept constant at $0.4$.
First this results in worsening - i.e. increasing - values of $K_{\rm p}^{-1}$ and thus of superconducting susceptibility and quasi-order for all three density-pairs $(n_{\rm p},n_{\rm m})$.
Second, it goes on until, between $U=2.0$ and $U=1.0$, a cross over from the pair-phase-limited to the amplitude-limited regime has occurred, as attested by the loss of the strictly monotonic decay of $K_{\rm p}^{-1}$ as $n_{\rm p}$ is decreased and $n_{\rm m}$ is increased.
At the same time, lowering $U$ also simultaneously \emph{lowers} the single-particle length $\xi_{S,{\rm m}}$ induced in the metal.
This behaviour indicates that even though the pairing in the P-subsystem is weakened as $U$ is lowered, overall the effective gap induced in the metal is still increased. 
This is explained by the fact that coupling between the two subsystems is deliberately chosen to be perturbative, which suppresses single-electron tunneling, and the leading process is a tunneling of pairs with an amplitude proportional to $t^2_\perp/(U-2(\mu_{\rm m}-\mu_{\rm p}))$.
Any version of Kivelson's proposal targets this regime, as unsuppressed single-electron tunneling between the subsystems is obviously detrimental to the performance of the entire setup.
The induced gap in the metal is thus generated by a sequence of virtual processes:
an on-site pair of electrons tunnels from the metal onto a site in the P-subsystem with amplitude $t^2_\perp/(U-2(\mu_{\rm m}-\mu_{\rm p}))$, lowers its energy by an amount on the order of $U-2(\mu_{\rm m}-\mu_{\rm p})$, and then tunnels back with that same amplitude.
As the $\xi_{S,{\rm m}}$-data in \cref{U4regime} confirms, the pair-tunneling, with an overall scaling of $t^4_\perp/(U-2(\mu_{\rm m}-\mu_{\rm p}))^2$ dominates the energy-lowering term scaling as $U-2(\mu_{\rm m}-\mu_{\rm p})$.
This explains the effective growing of the induced gap that is signaled by the lowering of $\xi_{S,{\rm m}}$ with $U$.

The behaviour of $K_{\rm m}^{-1}$ in regime~1 that is also shown in \cref{U4regime}, which characterizes the pair-pair correlations induced inside the metal by the P-subsystem (studied here for the first time) is distinctly different from that of $K_{\rm p}^{-1}$.
While the $K_{\rm m}^{-1}$-values are significantly improved over $K_{\rm m}^{-1}=1$ (i.e. compared to the isolated metal at $t_\perp=0$) they are primarily affected by the density imbalance between P- and M-subsystems.
In comparison, the impact of changes in $U$ on $K_{\rm m}^{-1}$ at any one of the three different pairs of $(n_{\rm p},n_{\rm m})$-values is far smaller, while that of changes in the relative density $n_{\rm p}-n_{\rm m}$ is far larger.
In this, $K_{\rm m}^{-1}$ behaves in an opposite manner to the {$K_{\rm p}^{-1}$}-data summarized in \cref{U4regime}. 
Our data in ref.\cite{JEEBOT2025} had indicated that both the ratio of the pairing energy of the P-subsystem to the kinetic energy of the metal, as well as the value of the coupling $t_\perp$, might play key roles in determining the superconducting susceptibility and pair-pair correlations overall.
In the present simplified systems, with no kinetic term in the P-subsystem, that ratio of energies is simply given by $U/t_{\rm m}$.
In \cref{U10regime3} (or \comm{\cref{fig:phase_diagram_tp3} in \cref{{app:thirdstate}}}) we show the comprehensive data on $K^{-1}_{\rm p,m}$ and $\xi_{S,{\rm m}}$ for regime~2, in which $U$ is fixed to $U=10.0$, and $t_{\rm m}$ and $t_\perp$ are varied.
Results are given for all three density-pairs $(n_{\rm p},n_{\rm m})$ considered in this work, for $t_\perp=3.0$ in the top row, and for $t_\perp=2.0$ in the bottom row, one separate graph for every value of $t_{\rm m}$, from $t_{\rm m}=10.0$ to $t_{\rm m}=7.0$.
To ease comparison with the results of ref.~\cite{JEEBOT2025}, the densities have been converted to nominal Fermi-vectors, ${k^\lambda_{{\rm F}}=\pi n_\lambda /2}$ (top and bottom $x$-axis of each subfigure).

\begin{figure}[ht!]
    \centering
     \scalebox{0.8}{
 \includegraphics[scale=1]{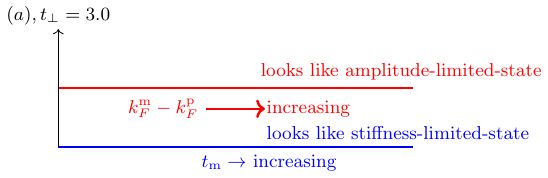}
    }
        \scalebox{0.8}{
    \includegraphics[scale=1]{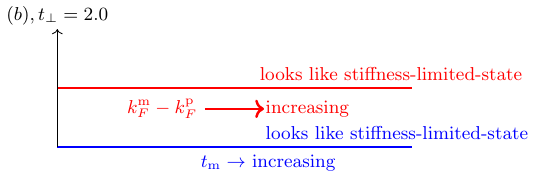}
    }
    \caption{
    \comm{(a) The third intermediate state (b) expected behavior according to~\cite{JEEBOT2025}, with $k^{\rm m}_{F} > k^{\rm p}_{F}$.}
    }
    \label{fig:phase_draw}
\end{figure}

\begin{figure*}[t!]
            \centering

        %
\includegraphics[scale=1]{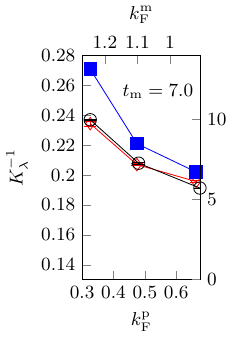}
        %
        %
\includegraphics[scale=1]{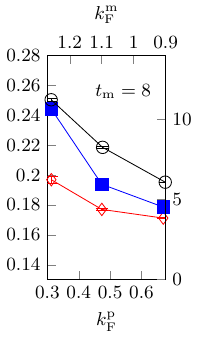}
        %
        \includegraphics[scale=1]{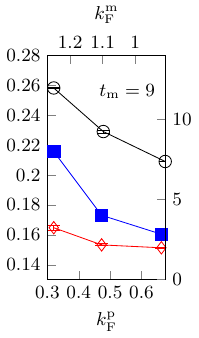}
        %
        %
        \includegraphics[scale=1]{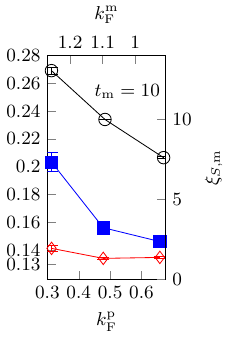}
        %
%
%
        %
        %
        \includegraphics[scale=1]{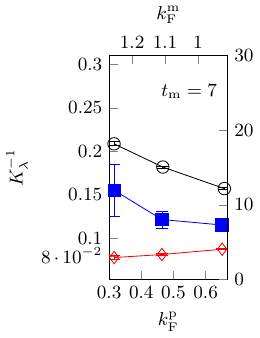}
        %
        \includegraphics[scale=1]{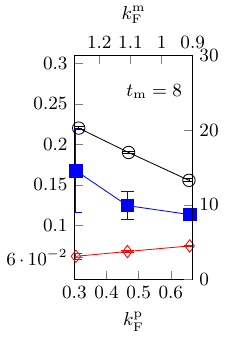}
        %
         \includegraphics[scale=1]{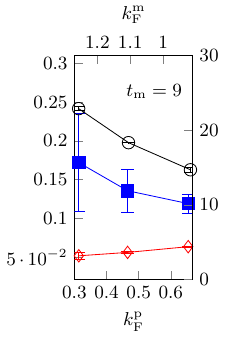} 
        %
        %
        %
         \includegraphics[scale=1]{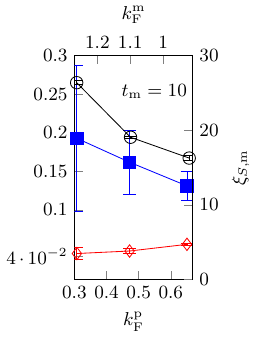}
\caption{
        (Left $y$-axis) Luttinger liquid parameters in the P-subsystem ($K^{-1}_{\rm p}$, \protect\diamondline) and M-subsystem ($K^{-1}_{\rm m}$,  {{\protect\squareline}}) versus $k^{\rm p}_{F}$, and  (right $y$-axis) the single-particle length scale $\xi_{S,\rm m}$ (\protect\opencircleline) against $k^{\rm m}_{F}$ in  regime~2 with $L=200$,
         $t_{\perp}=3.0$ (top row) and  $t_{\perp}=2.0$ (bottom row).
        The corresponding P- and M-subsystem wave-vectors $k^\lambda_{\rm F}$ can be read off from the bottom $x$-axis and top $x$-axis respectively.
        }
    \label{U10regime3}
\end{figure*}

Several key messages about the physics of these systems emerge out of these results.
Comparing the behaviour of $K_{\rm p}^{-1}$ with decreasing $k^{{\rm p}}_{{\rm F}}$ (corresponding to an increase in the detuning of the densities in the two subsystems) in the top and bottom rows reveals that changing $t_\perp$ can amount to changing whether the P-subsystem is in the stiffness-limited regime, the amplitude-limited state, or in a third, intermediate, state.
Assessing the top row, with $t_\perp=3.0$, comprehensively confirms the existence of this third state, where the P-subsystem seems to show behavior that is partially in line with an amplitude-limited and partially in line with a stiffness-limited state\comm{, as summarized in the top figure of \cref{fig:phase_draw}.}
When the value of $t_{\rm m}$ is fixed, the system shows behaviour consistent with the typical amplitude-limited state that we found in ref.~\cite{JEEBOT2025}.
%
%
%
%
%
%
%
Increasing the difference $|n_{\rm p}-n_{\rm m}|$ weakens the gap induced in the metal by the proximity effect emanating from the P-subsystem, as attested by the growth of $\xi_{S,{\rm m}}$.
This in turn leads to a strengthening of the reverse proximity effect that the metal exerts on the P-subsystem, whereby its intrinsic pairing is weakened by the additional kinetic energy available for electrons performing virtual tunneling into the metal.
When the P-subsystem has little pairing energy to spare - which is the effective definition of the amplitude-limited state - any boost to the reverse proximity effect leads to an overall degrading of the superconducting susceptibility and pair-pair-correlations of the P-subsystem.
This is attested by the worsening - i.e. growing - values of $K^{-1}_{\rm m}$ as $|n_{\rm p}-n_{\rm m}|$ (and thus $|k^p_{\rm F}-k^m_{\rm F}|$) is increased.
However, comparing data for \emph{different} values of $t_{\rm m}$ with each other in the top row of \cref{U10regime3} reveals an effect that had not previously been discussed. 
Namely, the pair-pair correlations induced inside the metal, which are characterized by $K^{-1}_{\rm m}$ (also shown in \cref{U10regime3}), can feed back into the P-subsystem and significantly mitigate the negative impact of detuning $n_{\rm p}$ from $n_{\rm m}$.
Specifically, $K^{-1}_{\rm m}$ is enhanced comprehensively - i.e. lowered - as $t_{\rm m}$ is raised. 
As the data demonstrates, this enhancement of $K^{-1}_{\rm m}$ goes hand in hand with a significant abatement of the amplitude-limited behavior of $K^{-1}_{\rm p}$ as $k^p_{\rm F}$ is detuned from $k^m_{\rm F}$.
The quasi-superconducting mode induced inside the metal can thus partially support the superconducting properties of the P-subsystem.
This finding illuminates the fact that the metal may actually mediate pair-pair coupling within the P-subsystem in two ways:
The first, as discussed above, consists of coherent but separate propagation of the constituent electrons of a pair inside the metal, across distances of scale $\xi_{S,{\rm m}}$, and thus above the proximity-effect-induced gap.
But as we see here, a pair of electrons entering the metal from the P-subsystem may also propagate coherently \emph{as a pair} inside the metal, i.e. below the induced gap, after which it may re-enter the P-subsystem, thereby also mediating effective ranged pair-pair coupling within the P-subsystem.
Whereas the first mechanism competes with and may thus weaken pairing within the P-subsystem, which becomes noticeable when the system is in a amplitude-limited state, the second one is compatible with the pairing.

However, the bottom row of \cref{U10regime3}, which collects the results for $t_\perp=2.0$, illustrates how contingent the effectiveness and even the underlying physics of this second channel for mediated pair-pair-coupling is on other parameters.
At this lower level of coupling between the P- and M-subsystems, $K^{-1}_{\rm m}$ is either not enhanced, or even worsened (i.e. increased) as $t_{\rm m}$ is increased.
At the same time $K^{-1}_{\rm p}$ is markedly improved at all parameters compared to the case of $t_\perp=3.0$.
And only the detuning of $k^p_{\rm F}$ $k^m_{\rm F}$, and thus of relative densities, appreciably lowers $K^{-1}_{\rm p}$ yet further - the core behaviour of P-subsystems in a state of low-pair-phase-stiffness state~\cite{JEEBOT2025,JEEBOT2025_3}:
in this state, the P-subsystem prioritizes extended-range pair-pair coupling mediated via the metal in order to achieve the highest possible superconducting susceptibilities, i.e. the lowest possible values of $K^{-1}_{\rm p}$.
This means prioritizing high $\xi_{S,\rm m}$ - the high intrinsic pairing of the P-subsystem in this state means that it can stand to loose some of that pairing energy due to larger $\xi_{S,\rm m}$-values, which is more than compensated for by the improved range of mediated pair-pair coupling that a larger $\xi_{S,\rm m}$ affords simultaneously.

\begin{figure}
\centering
    \includegraphics[scale=1]{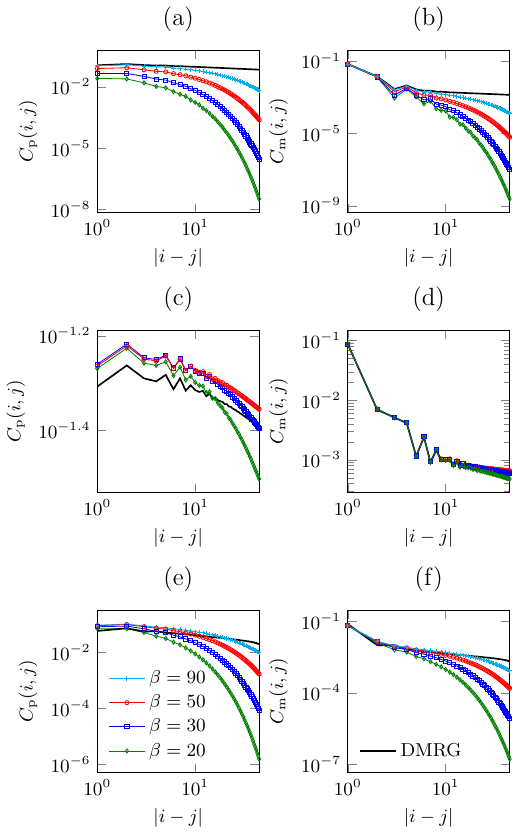}
    \caption{
    Finite temperature data: $C_{\rm p}(i,j)$ (left column) and $C_{\rm m}(i,j)$ (right column)  in the three different density regimes with $L=200$.
     [(a) and (b)] regime~1 at $U=1.0$, $n_{\rm p}=0.51$ and $n_{\rm m}=0.65$
     [(c) and (d)] regime~2 at {$t_{\rm m}=10.0$, $t_\perp=3.0$}, {$n_{\rm p}=0.2$ and $n_{\rm m}=0.8$}
     [(e) and (f)] regime~3 with $t_\perp=1.0$, $n_{\rm p}=0.51$ and $n_{\rm m}=0.65$.
    $\beta=90$ ({\color{cyan}$+$}),$\beta=50$ (circle, \protect\opencircle), $\beta=30$ (square, \protect\opensquare),$\beta=20$ (diamond, \protect\opendiamond).  
    %
    \comm{The lines without marker are the corresponding DMRG results. [(e) and (f)] $L=100$ for the DMRG results.}
    }
    \label{fig:afqmc}
\end{figure}
Finally, we have used AFQMC-based numerics to investigate the behaviour of these systems at finite temperature.
%
{\cref{fig:afqmc} shows one example from each of the three regimes,} once the temperature has dropped to around $0.2$\% of $U$, the near-long-ranged superconducting correlations $C_\lambda(i,j)$ are virtually indistinguishable from those of the ground state in both the P- and M-subsystems.
This is true even for the largest systems that we have simulated ($L=200$).
\section{Implications for Anderson- and Kondo-lattices realized in heavy-fermion materials}
\label{sec:impl}
We had previously established that an exact mapping between Kivelson's proposal and the periodic Anderson- and Kondo-lattices exists~\cite{JEEBOT2025_3}, as re-stated in \cref{sec:setup}. 
We now discuss the implications of the present results towards the physics of those lattice-models.
And while the present work studies the 1D version, and strictly speaking our results only apply in that context, the mapping is generic to bipartite lattices in arbitrary dimensions.
We may thus at least attempt some general inferences regarding higher-dimensional models as well.

First off, being doped away from unit filling maps to the presence of a magnetic field along the $z$-axis for the Anderson-lattice model in \cref{al_ham}, or for the Kondo-lattice model that is commonly derived from it in the limit of weak coupling $t_\perp/U\ll 1$ (cf. ref.~\cite{JEEBOT2025_3}).
Without such a magnetic field, we had previously found that the 1D versions of these systems appear to have near-ordered antiferromagnetic (AFM) correlations that are degenerate along all three spin-axes, but that these will always be exponentially suppressed in the thermodynamic regime even at zero temperature due to the unavoidable presence of a spin gap~\cite{JEEBOT2025_3}.
The degeneracy of the AFM-correlations corresponded to the degeneracy of the superconducting and density-density correlations at unit filling, $n=1$, for Hamiltonian \cref{model_equ}.
Now, in the presence of a magnetic field, i.e., when being partially spin-polarized, Hamiltonian \cref{al_ham} will exhibit a clear breaking of the degeneracy between the AFM correlations.
Those in the $x-y$ plane will be significantly enhanced over spin-spin correlations in the $z$-direction, which will appear suppressed in comparison.
This follows directly from the breaking of the degeneracy between superconducting and density-density correlations in the original model \cref{model_equ}, as exemplified by the data shown in \cref{fig:nocross_over} and \cref{fig:U4pair-pair_corr}, in conjunction with \cref{eq:corr_trafo}.
Furthermore, as we had already discussed in ref.~\cite{JEEBOT2025_3}, in the present work we find as well that the much-studied RKKY-interactions between the localized spins mediated via the metal are not long-ranged (i.e. low-exponent algebraically decaying) at all.
That is what perturbative analytical treatments that approximate the dense lattice of spins as two dilute ones predict it to be~\cite{Rusin2017,Nejati2017}.
Instead, the exponential decay of the single-particle correlations $S_{\rm m}(i,j)$, as characterized by $\xi_{S,{\rm m}}$, signals that the spin-spin coupling mediated by the metal, decays exponentially.

The near-long-range ordering exhibited by the superconducting correlations analyzed in \cref{fig:nocross_over} - \cref{fig:afqmc} can thus be translated into an immediate prediction that any realization of a 1D Anderson- or Kondo-lattice model inside an external magnetic field, such as e.g. in ultra cold atomic gases, engineered chains of ad-atoms~\cite{Neel2011} or quasi-1D HFMs such as CeCo$_2$Ga$_8$ ~\cite{Wang2017}, can be marked by strong easy-plane AFM correlations for the spins in the localized orbitals, while spin-correlations along the $z$-axis will be much weaker.
Different from the half-filled case, these systems will \emph{not} be spin-gapped (cf.~\cite{JEEBOT2025_3} and references therein) and thus will not be paramagnetic.
If $K^{-1}_{\rm p}$ is especially small, and the realized system is not too large, the localized spins may even appear to enter an ordered easy-plane AFM state.
As to the implications of our results for 2D and 3D Anderson- and Kondo-lattices, as realized in a range of HFMs: 
without an external magnetic, the ground state of the Kondo-lattice does not appear to be precluded from realizing true AFM-order below a critical coupling between the localized spins and the conduction band~\cite{Assaad1999}, the way it is for the 1D-version of this problem (cf. above and ref.~\cite{JEEBOT2025_3}).
Based on the known physics of the HFMs, the same is true in 3D~\cite{Stewart1984}.
Our findings in the present work thus make it very likely that beyond some critical magnetic field the 2D and 3D versions of \cref{al_ham} will realize true easy-plane AFM-order, in the $T=0$ ground state in 2D, at finite temperature in 3D.
As part of the existing efforts at studying partially spin-polarized HFMs~\cite{Aoki2013}, such a transition may already have been found under increasing external magnetic fields~\cite{Palacio_Morales2015,Pourret2017}.
Using the inverse of the exact particle-hole mapping that we used to derive \cref{al_ham} from \cref{model_equ} thus opens the possibility for turning partially spin-polarized HFMs into simulators for many different variants of Kivelson's proposal. 
In such a set-up, any occurrence and characterization of an easy-plane AFM-order could be mapped exactly to superconductivity driven by coupling a P-subsystem (i.e. the spins occupying the localized orbitals) to a metallic reservoir (i.e. the itinerant conduction band of the HFM).

\section{Discussion and conclusions}\label{sec:disc}
In the present work, we have looked at the simplest possible variant of Kivelson's proposal for enhancing superconducting properties via a metallic bath that is amenable to large-scale and quasi-exact many-body calculations approaching the thermodynamic limit.
Comprised of a chain of disconnected negative-$U$ sites, the P-subsystem, and a 1D metallic chain, we show that when doped well away from half-filling there is no sign of a charge-gap, which is known to emerge and destroy even quasi-order at half-filling.
From there, we show that these systems can support near-long-range superconducting quasi-order within the P-subsystem, in which the associated correlations appear to decay only barely within the bulk of the system, across a broad range of Hamiltonian-parameters.
The ground-state calculations via DMRG reveal the complex manner in which the two subsystem mutually affect each other, as well as the system's superconducting properties.
In particular, they reveal the presence of two channels whereby the metal creates or enhances superconducting correlations  within the P-subsystem.
The first of these is the coherent but separate motion of electrons through the metal, taking place above the spin gap that the proximity-effect emanating from the P-subsystem.
\comm{The second mechanism consists of electrons continuing propagating \textit{as pairs} in the M-subsystem, which they can do due to the proximity-effect-induced gap.}
At the same time, we find density-density correlations to be suppressed significantly, which had been degenerate with the superconducting ones at half-filling.

\section*{Acknowledgements}\label{sec:acknow}
This work was supported by an ERC Starting Grant from the European Union’s Horizon 2020 research and innovation programme under grant agreement No. 758935; and the UK’s Engineering and Physical Sciences Research Council [EPSRC; grant number EP/W022982/1 and UKRI2088]. 
This work is also supported by the Swiss National Science Foundation under grant number 200020-219400.
The computations were enabled by resources provided through multiple EPSRC ``Access to HPC'' calls (Spring 2023, Autumn 2023, Spring 2024 and Autumn 2024) on the ARCHER2, Peta4-Skylake and Cirrus compute clusters, as well as by computer time awarded by the National Academic Infrastructure for Supercomputing in Sweden (NAISS).
This work was supported by a grant from the Swiss National Supercomputing Centre (CSCS) under project ID s1307 on Alps.
We further acknowledge the EuroHPC Joint Undertaking for awarding this project access to the EuroHPC supercomputer LUMI, hosted by CSC (Finland) and the LUMI consortium through a EuroHPC Regular Access call.
The authors gratefully acknowledge the HPC RIVR consortium and EuroHPC JU  for funding this research by providing computing resources of the HPC system Vega at the Institute of Information Science of the Republic of Slovenia.
The authors also acknowledge the use of the HWU high-performance computing facility (DMOG) and associated support services in the completion of this work.

\appendix

\section{DMRG \comm{and AFQMC}}
\label[appendix]{app:DMRG}
We use the density matrix renormalization group (DMRG) method based on matrix product state (MPS) to calculate ground-state properties at $T=0$ of Eq. \ref{model_equ} with $\mu _{\rm p}=0$ \cite{Schollwock2011}. 
The calculations are done on the setup with an open boundary condition (OBC) and keeping up to 8000  bond dimension. 
At this bond dimension, the iteration is performed several times with a truncated weight of approximately $\mathcal{O}(10^{-14})-$ $\mathcal{O}(10^{-16})$. 
Throughout the numeric's, we have used two site updates and an energy truncation of about $10^{-14}$ to $10^{-16}$ \cite{JEEBOT2025}.
\comm{In contrast, the AFQMC calculations use periodic boundary conditions (PBCs).}
\comm{Throughout this work, all data analysis such as fitting of correlation functions of the form $\langle\hat{A}_i\hat{B}_j\rangle$, where $i$ is a site in the center of the system, is performed within the central half of the system, i.e. for ${L/4\leq j \leq 3L/4}$, in order remove boundary effects (i.e. Friedel oscillations) arising from the open boundary conditions, and thus approximate the properties of the bulk in the thermodynamic limit. 
Therefore, all quantities such as $K_\lambda^{-1}$, $\xi_{S,\rm m}$ etc. extracted from such fits and shown in this work include the resulting error estimates arising from such fits.}
\section{Regime~2 and the intermediate state}
\begin{figure}
\centering
\includegraphics[scale=1]{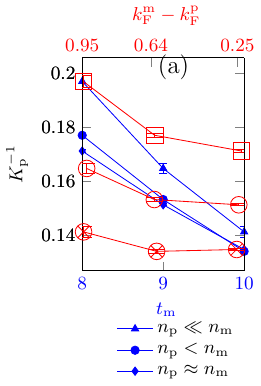}
\includegraphics[scale=1]{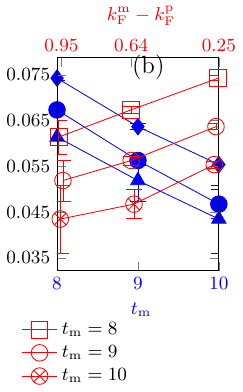}
    \caption{ Re-plot of \cref{U10regime3}. 
    (a) $t_\perp =3.0$ and (b) $t_\perp=2.0.$
    }
    \label{fig:phase_diagram_tp3}
\end{figure}

\begin{figure}[ht!]
    \centering
\includegraphics[scale=1]{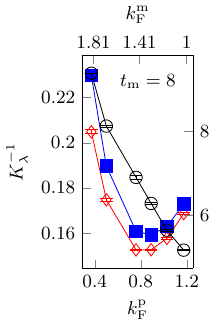}
\includegraphics[scale=1]{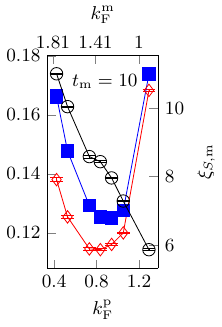}
\includegraphics[scale=1]{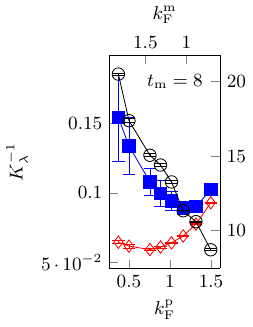}
\includegraphics[scale=1]{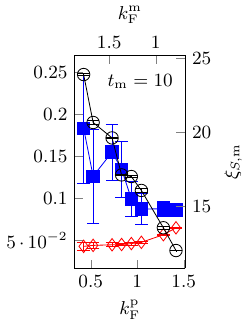}
    \caption{
    (Left $y$-axis) Luttinger liquid parameters in the P-subsystem ($K^{-1}_{\rm p}$, \protect\diamondline) and M-subsystem ($K^{-1}_{\rm m}$, {{\protect\squareline}}) versus $k^{\rm p}_{F}$, and  (right $y$-axis) the single-particle length scale $\xi_{S,\rm m}$ (\protect\opencircleline) against $k^{\rm m}_{F}$ in  regime~2 with $L=200,t_\perp=3.0$ (top row), $t_\perp=2.0$ (bottom row), and a global density of $0.7$.
        %
        %
        The corresponding P- and M-subsystem wave-vectors $k^\lambda_{\rm F}$ can be read off from the bottom $x$-axis and top $x$-axis respectively.
    }
    \label{fig:n07}
\end{figure}

\label[appendix]{app:thirdstate}
In this section, we show additional results in regime~2 with a global density of $n=0.7$ and for $t_{\rm m}=10$ and $8$ only. 
As depicted in~\cref{fig:n07}, the system is in the stiffness-limited regime at $t_\perp=2.0$ (bottom row) when the difference $k_{\rm m}-k_{\rm p}$ or $t_{\rm m}$ increases.
At $t_\perp=3.0$ (top row), the system is in a stiffness-limited state when $t_{\rm m}$ is increasing  and in an amplitude-limited state when the difference between $k_{\rm m}$ and $k_{\rm p}$ increases ($k_{\rm m} >k_{\rm p}$).
Consequently, the $t_\perp=3.0$ parameter space again is in an intermediate state where the system is partially in a stiffness-limited and partially in an amplitude-limited state.
In conclusion, this means that this system shows similar physical behavior at these different global densities (including at $n=0.8$-data not shown here), see~\cref{U10regime3},~\ref{fig:phase_diagram_tp3} and~\ref{fig:n07},
%
%
thus, pointing to the fact that the physics in this system does not depend on the global density but rather on the inter-layer densities.
Also, we note that calculations for higher global densities, such as 0.9, 0.95, etc. at $L=200$ are quite difficult, as there are stuck in metastable excited
state and one might need higher bond dimensions than in~\cref{app:DMRG}.

%
%
%
%
\FloatBarrier
\bibliography{ref}
\end{document}